\documentclass[a4paper,10pt]{article}
\usepackage[english]{babel}
\usepackage{graphicx}
\usepackage{epsfig,epstopdf}
\usepackage{subfigure}
\usepackage{amsmath}
\usepackage{latexsym,amssymb}

\usepackage[usenames]{color}
\newcommand{\crz}{\color{Black} } 

\newtheorem{proposition}{Proposition}

\def\Re{\mathbb{R}} 

\def\P{\textit{P\/}}
\def\RP{\textit{RP\/}}
\def\DP{\textit{DP\/}}
\def\LRP{\textit{LRP}}

\title{
Covering a line segment with variable radius discs }
\author{Alessandro Agnetis\thanks{Universit\`{a} di Siena, Dipartimento di Ingegneria dell'Informazione, via Roma 56, 53100
Siena,
Italy.} \and Enrico Grande\thanks{ATLAS Center, The University of
Arizona, Tucson AZ 85721. Email: pitu@sie.arizona.edu. Fax: 520-621
6555.} \and Pitu B. Mirchandani$^\dag$ \and Andrea
Pacifici\thanks{Universit\`{a} di Roma ``Tor Vergata'', Dipartimento
di Ingegneria dell'Impresa, via del Politecnico 1, 00133 Roma,
Italy.}}

\begin{document}

\maketitle

\begin{center}
\begin{minipage}{360pt}
\begin{center}\textbf{Abstract}\end{center}

The paper addresses the problem of locating sensors with a circular
field of view so that a given line segment is under full
surveillance, {\crz which is termed as the \emph{Disc Covering
Problem on a Line}}. The cost of each sensor includes a fixed
component $f$, and a variable component $b$ that is proportional to
the field-of-view area. When only one type of sensor {\crz or, in
general, one type of disc,} is available, then a simple polynomial
algorithm solves the problem. When there are different types of
sensors in terms of the $f$ and $b$ parameters, the problem becomes
hard. A branch-and-bound algorithm as well as an efficient heuristic
are developed. The heuristic very often obtains the optimal solution
as shown in extensive computational testing.

\par\vskip0.2cm\noindent

\begin{center}\textbf{Scope and Purpose}\end{center}
Problems of locating facilities to cover sets of points on networks
and planes have been widely studied. This paper focuses on a new
covering problem that is motivated by an application where a line
segment  is to be kept under
surveillance using different types of radars. Using reasonable
assumptions, some nonlinear covering problems are formulated.
Efficient exact algorithms and heuristics are developed and analyzed
for ``easy'' and ``hard'' cases, respectively.
\par\vskip0.5cm\noindent
\textbf{Keywords:} Sensor location, network covering problems, mixed
integer nonlinear programming.
\end{minipage}
\end{center}

\section{Introduction}\label{sec:intro}
In this paper we introduce and study a new locational decision
problem: given a set of {\crz discs with} variable {\crz radii} with
costs depending on their radii and {\crz fixed costs}, find a subset
covering a unit length segment at minimum cost.

This problem {\crz was} motivated by the following application, part
of which {\crz was an industry-funded radar surveillance project at
The} University of Arizona.
We have a river over which we need to track possible activities of
non-collaborative or antagonistic objects {\crz or people }(e.g.,
unauthorized boats, dangerous floating objects, {\crz swimmers,
etc}). For this purpose, we need to locate a set of radars so that
every point on the river is under surveillance by at least one
radar.
It is assumed that the river can be modeled as a tree network
consisting of line segments and that {\crz each} radar has a field
of view defined by a radius and an angle of view (a pie-shaped
coverage), with this angle large enough so that the coverage area
may be approximated as a disc. Although the problem is relatively
easily stated, the actual locational decision is complicated {\crz
due to} several additional factors. {\crz Coverage} depends not only
on the river topology, radar type and power, but also on several
parameters such as width of {\crz river} and obstacles over {\crz
it}, potentially forbidden {\crz areas where radars may not be
located}, elevation of the {\crz potential} location sites, and
other characteristics associated with the physical environment,
dealing with, for example, the atmospheric and water conditions.
Further details on {\crz this scenario} and the scope of {\crz the
project} are reported in~\cite{techrep}.

This radar sensors location model relates to several broad classes
of geometric locational problems.
Many important land-use planning decisions deal with locating
facilities at sites, choosing from a given set of potential sites,
so that another given set of points are {\crz ``}covered{\crz ''
(i.e., they are within a specified distance from the closest located
facility) while optimizing} a specified objective. {\crz Models for
locating at points within continuous spaces, as well as locating
among set of discrete points or on a network, are} widely used by
geographers, regional scientists, {\crz network planners, and
others} facing {\crz locational} decisions problems {\crz  which can
be modeled as such covering problems} (for a comprehensive review of
the literature see, for example,
\cite{ghosh-rushton,kolen-tamir,love-et-al.,pitu1}). {\crz From} the
methodological {\crz viewpoint}, {\crz the radars location problem
relates closely} to the class of geometric covering problems where
potential facilities and demand points are embedded {\crz on a}
Euclidean plane, {\crz for which there is considerable literature}.
{\crz We briefly review below results that are most relevant for our
application.}

{\crz Problems related to \emph{Covering with discs}} consists of
identifying the minimum number of discs with fixed radius to cover a
given set of points in the plane. A number of articles {\crz have}
appeared in the last {\crz three} decades {\crz addressing} this
NP-hard problem. In 1975, Chv\'{a}tal introduced the \emph{Art
Gallery Problem} in \cite{chvatal}, where one has to find the
minimum number of watchmen ({\crz or} cameras) {\crz needed} to
{\crz observe} every {\crz wall} of an art gallery room. The art
gallery is assumed to be a $n$-sided polygon, possibly with
polygonal holes. It has been shown that an art gallery with $h$
holes and $n$ walls (including holes' sides) requires at most
$\lfloor (n+h)/3\rfloor$ watchmen (the bound is tight,
see~\cite{hoffmann,orourke87}). Another {\crz important} paper, by
Hochbaum and Maas \cite{hoch-maas}, presents polynomial
approximation algorithms for different versions of geometric
covering {\crz problems}, including covering by discs. {\crz
Subsequently, several} papers {\crz have} appeared {\crz with}
improved approximation factors and running times (see for {\crz
example}, \cite{bron95,franceschettietal01}).

{\crz The problem
of partial covering}, also referred to as the robust $k$-center
problem, is analyzed {\crz in~\cite{xiao03}, where computational}
complexity is discussed and approximation algorithms together with
{\crz computational} evidence of their performance are provided.

{\crz The geometric disc covering problem relates also to the
deployment of} wireless transmission networks. Surveys on coverage
problems dealing with this particular application can be found
in~\cite{huangetal05} and~\cite{wang06}. We limit {\crz our
literature review to} a few papers dealing with applications similar
to the radar sensors location {\crz problem}.
{\crz Alt \emph{et al.}~\cite{altetal06} consider a problem where a
set of points demand connectivity}. {\crz Their goal} is to locate a
set of sensors, modeled as discs with variable radii, covering the
demand points at minimum total cost. Each sensor's transmission cost
has the form $f(r) = r^\alpha$ where $r$ is the covering radius of
the {\crz sensor}. Several results are {\crz presented
in~\cite{altetal06}}, including complexity characterization and
approximation algorithms. Although different scenarios are
addressed, depending on possible restrictions on discs' locations
and demand points, {\crz their} analysis is limited to discrete sets
of points.

{\crz Article~\cite{galotaetal01} addresses the problem of locating base stations
for wireless communication where the demands and potential facilities are represented by a
discrete set of points and each station can broadcast up to a maximum distance.}
A polynomial approximation scheme is given, together with
{\crz complexity} results.
{\crz The following disc-covering geometric problem applied to
wireless communication is addressed by Franceschetti et
al.~\cite{franceschettietal04}:} given an infinite square grid $G$,
determine how many discs, centered at the vertices of $G$, with a
given radius $r$, are required, in the worst case, to completely
cover a disc with the same radius arbitrarily placed on the plane.
The authors show that this number is an integer in $\{3,4,5,6\}$
depending on $r$ and on the grid spacing. {\crz In addition,} they
discuss the applicability of this model to the design of
approximation algorithms for facility locations on regular grids and
to base station placement {\crz for wireless communication}.
The expected quality of service (level of surveillance) of a given sensor network is analyzed in~\cite{meguerdichianetal01} and~\cite{xiangetal03}, {\crz where the authors exploit} computational
geometry and graph theoretic techniques, such as Voronoi diagrams, Delaunay triangulation and graph search, to design exact polynomial algorithms for some special cases.

{\crz Location of railway stops} is another application of the disc covering
problem. In~\cite{hamacheretal01}, the effect of
introducing additional stops in the existing railway network is addressed. The problem is comprised of covering a set of points in the plane by discs with the restriction that their centers have to lie on a set of {\crz line segments that represents the railway tracks}. A similar problem is addressed
in~\cite{mammanaetal03}, where the discs must be centered on two intersecting lines.

The location problem we address in this paper is a special disc covering problem in the following ways:
\begin{itemize}
\item There are different types of facilities, which in our case are
      radar sensors, where the area covered by each radar is a disc
      with a diameter $x$ that depends on the power of the radar unit.
\item The cost $c_i$ of locating {\crz disc} $i$ includes a nonnegative fixed cost $f_i$
      and a variable cost, which may be approximated by an homogeneous
polynomial function $g(x_i)$. In particular, $g(x_{i})$ is modeled
as a second-order polynomial.
\item {\crz A line segment with negligible width has to be covered by the discs}. As it will be clear in Section~\ref{sec:model}, we may assume that the segment has unit length with no loss of generality.
\end{itemize}
We refer to our problem as the \emph{Disc Covering Problem on a
Line.}

The paper is organized as follows. In Section~\ref{sec:notation}
some preliminary results {\crz are presented for the case of identical disc (radar)} types and convex cost
functions. In
Section~\ref{sec:model}, the general case is presented and a
quadratic programming formulation is {\crz developed}. A Lagrangean relaxation
of the problem and a technique to solve such a relaxation is also
proposed. Section~\ref{sec:bounds&algo} presents a branch-and-bound
algorithm for the problem: upper and lower bounding techniques are
illustrated and a branching strategy is discussed. Some
computational results are given in Section~\ref{sec:exp}. Finally,
some concluding remarks are made in Section~\ref{sec:concl}.

\section{Notation and preliminary results}\label{sec:notation}

We denote by $Q$ the set of the $q$ available {\crz discs (radars)}.
{\crz For all} $i\in Q$, \emph{at most one copy} of {\crz disc} $i$  
may be used for covering the line segment and any power level is
allowed {\crz so that} we can have any {\crz disc} coverage distance
$0 \leq x_i \leq 1$. These assumptions may appear {\crz restrictive
for  real applications but note that} $(i)$ usage of multiple copies
of the same {\crz disc} type may be modeled by including in $Q$ a
suitable number of items with the same characteristics and $(ii)$ if
{\crz a limit $D$ exists on the coverage distance, then the} problem
may be decomposed by splitting the segment {\crz into pieces whose
lengths are} not greater than $D$ and solving the problems for each
segment {\crz separately} (this may be an effective heuristic
approach).

For any selected {\crz disc} $i\in Q$, the coverage {\crz distance}
is the diameter of the disc $x_{i} \in \mathbb{R}_{+}$, and its
contribution in the total cost function is
$$c_i(x_i) = \left\{
\begin{array}{ll}
  0 & \mbox{if}\ x_{i} = 0\\
  f_{i} + g_i(x_{i}) & \mbox{if}\ x_{i}>0
\end{array}
\right.,$$
where $g_i(\cdot)$ is convex with $g_i(0) = 0$ and the \emph{setup
cost} $f_i$ is nonnegative. Although, {\crz because of the fixed
cost component}, the cost function $c_i(x_{i})$ is nonconvex in $0
\leq x_i \leq 1$, {\crz when the set of selected discs $S$, i.e.,
those for which $x_i > 0$, is fixed, then total coverage is in fact
convex} and the problem of determining the {\crz covering} diameters
is easily solved {\crz using} KKT conditions (see
Section~\ref{sec:model}).

\begin{figure}[htbp]
\begin{center}
\includegraphics[scale=1,angle=0]{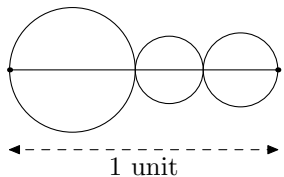}
\end{center}
\caption{\small Example of a feasible solution}
\label{fig:feasiblesol}
\end{figure}

Note that once $x_i$ is given for all $i\in Q$ (we will have $x_i =
0$ for those radars that are not selected), {\crz it is trivial to
find the set of optimal locations:} just align the discs so that
they do not intersect and {\crz they} cover the entire line. For
this purpose we choose the diameters in such a way that their sum is
equal to the length of the line, {\crz which in our case} is equal
to 1. An {\crz illustration} of a feasible solution to our problem
is given in Figure~\ref{fig:feasiblesol}.

{\crz We now} present some simple results concerning the case when
all available {\crz discs} are of the same type, that is, for all
$i\in Q$ and {\crz $x > 0$}, $c_i(x)=c(x)$ is a general nonnegative
convex function.
%
To the best of our knowledge, these results, though straightforward,
are not present in {\crz the} literature. However, it is worthwhile
to point out that, differently from~\cite{altetal06}, where the
objective function is of the form  $r^\alpha$ and both the potential
facility locations and the demand points are discrete sets, we
exploit the fact that we deal with a {continuous} line segment to obtain efficient solutions for even more
general cost functions.

{\crz When the discs are all of the same type, our} problem reduces
to finding the optimal number $k \leq q$ of copies and the optimal
coverage area for each copy.

\begin{proposition}\label{prop:equaldiamisbetter}
{\crz When all the discs} are of the same type having the cost
functions $c_i(\cdot) = c(\cdot)$, for all $i\in Q$, if an optimal
solution consists of locating $k$ {\crz discs}, then there is one
solution {\crz where each of the $k$ discs has the same diameter}.
\end{proposition}

\noindent\textbf{Proof.} Based on the convexity of $c(x)$, for any
$k$-uple of nonnegative numbers $x_{1}, \ldots, x_{k}$, with
$\sum_{i=1}^{k}{x_{i}}=1$, we have:
\begin{equation}\label{eq:equaldiamisbetter}
k\cdot c\left(\frac{1}{k}\right)\leq \sum_{i=1}^{k}{c(x_{i}).}
\end{equation}
Hence, the {\crz cost} of locating $k$ {\crz discs} (of the same
type) with equal diameters---that is, each {\crz disc} covers {\crz
an equal} portion of the line segment---does not exceed the cost of
any other feasible solution that uses $k$ {\crz discs}.
$\blacksquare$

{\crz Proposition}~\ref{prop:equaldiamisbetter} clearly indicates
the {\crz optimal locations of the discs} since they need to be
uniformly spaced over the line segment.

The next natural question we need to ask is ``What is an optimal
number of such {\crz discs, that is the best value for k?''}.

First we write $c(x) = f + g(x)$ with $f \geq 0$ and $g(0) = 0$.
Note that we may install at most $q$ {\crz discs} of the same type
on the line. It is easy to observe that with {\crz zero} setup costs
($f = 0$) the cheapest solution consists of installing the largest
possible number ($q$) of facilities.

A solution that uses $k+1$ {\crz discs} costs no more than a
solution with $k$ {\crz discs} if and only if the following is true:
\begin{equation}\label{eq:(k+1)radars_setup>0}
f + (k+1)\cdot g\left(\frac{1}{k+1}\right) \leq k\cdot g\left(
\frac{1}{k}\right).
\end{equation}
If $f=0$, {\crz the last inequality is always valid, because of the
convexity of $g(\cdot)$}. Therefore, it is cost-effective to locate
another {\crz disc} if the additional setup cost does not exceed the
gain in the variable costs.

The {\crz effective} cost of locating $k$ {\crz discs} is
\begin{equation}\label{eq:costtolocatekradars}
F(k) = kc(\frac{1}{k}) = kg\left(\frac{1}{k}\right) + kf.
\end{equation}
Since $g(\cdot)$ is convex,
\begin{equation}\label{eq:g()isconvex}
\left.\frac{\partial^{2}F}{\partial
k^{2}}=\left(\frac{1}{k^{3}}\right)\frac{\partial^{2}g}{\partial
k^{2}}\right|_{\frac{1}{k}} \geq 0
\end{equation}
Hence, there must be $1\leq k^{*}\leq q$ such that
\begin{equation}\label{eq:existsk*}
F(1)\geq F(2)\geq\ldots\geq F(k^{*}) \mbox{\ and}\ F(k^{*})\leq
F(k^{*}+1)\leq\ldots\leq F(q).
\end{equation}
Therefore, since a binary search can be used to efficiently find the
$k^{*}$, the following proposition holds.
\begin{proposition}
When the $q$ {\crz discs} are all of the same type with the cost
functions $c_i(\cdot) = c(\cdot)$, for all $i\in Q$, the problem is
solvable in $O(C\log(q))$ time, where ${C}$ is the maximum
computational effort for calculating $c(\frac{1}{k})$.
$\blacksquare$
\end{proposition}

\section{Problem formulation}\label{sec:model}
In this section we {\crz develop} a quadratic {\crz programming}
formulation of the general problem where the $q$ {\crz discs
(radars)} may have different properties.
%
%
Using the notation introduced in Section~\ref{sec:notation}, the
cost contribution of any {\crz disc} $i\in Q$ that covers an area
having diameter $x_i$ is a quadratic polynomial
$$c_{i}(x_{i})=f_{i} + b_{i}x_{i}^{2}$$
with $f_{i}\geq 0$, $b_{i}> 0$. {\crz Then the location} problem can be {\crz
formulated as} the following \emph{Mixed Integer Quadratic Program}.
\begin{center}
\begin{tabular}{cc}
(\P) & \begin{tabular}{r l r}
        $z^{*}=\min$ & $ \displaystyle \sum_{i \in Q}f_{i}y_{i} + b_{i}x_{i}^2$ &\\
        s.t.
        & $\displaystyle x_{i}\leq y_{i}$, for all $i\in Q$   & ($c1$)\\
        & $\displaystyle \sum_{i \in Q}x_{i} = 1$ & ($c2$)\\
        & $x \in \mathbb{R}_{+}^{q}$              & ($c3$)\\
        & $y \in \{0,1\}^{q} $                    & ($c4$)\\
        \end{tabular}
\end{tabular}
\end{center}
In the solution of \P\ , $y$ is the vector to indicate {\crz
selected discs (radars) in} $Q$ where $y_{i}=1$ if {\crz disc} $i$
is used, $y_{i}=0$ otherwise. Constraints $(c1)$ force the {\crz
disc} coverage diameter $x_i$ to be zero when the corresponding
{\crz disc} $i$ is not {\crz selected} (and therefore the
corresponding cost contribution is zero). Constraint $(c2)$ is the
\emph{coverage constraint} {\crz that assures that the whole line
segment is covered}.

As stated before, the unit length assumption does not introduce any
loss of generality. It is {\crz clear} that \P\ can be equivalently
used for a problem $\widetilde{\mbox{\P}}$\ where the line has
length $\ell\neq 1$. Let $\tilde{I}$ be an instance of
$\widetilde{\P}$ where the cost coefficients {\crz for the disc} $i$
are $\tilde{f}_{i}$ and $\tilde{b}_{i}$. Then we can solve
$\tilde{I}$ by solving an equivalent instance $I$ of the unit length
problem \P\ having cost coefficients $f_{i} = \tilde{f}_{i}$ and
$b_{i} = \ell^{2}\tilde{b}_{i}$. If $(x,y)$ is an optimal solution
of $I$, then $(\ell x,y)$ is optimal for $\tilde{I}$.


In the remainder of the paper we propose methods to solve problem
\P\ and discuss the results of some computational experiments to
evaluate the performance of these methods.

Our first observation concerns the existence of efficient methods to
find the optimal coverage when the set $S\subseteq Q$ of {\crz
selected discs (i.e. active radars)} is given or known \emph{a
priori}. Under this assumption, the variables $y_i = 1$ for all
$i\in S$ in problem \P\ and the resulting problem is easily solved
by applying Karush-Kuhn-Tucker optimality (KKT) conditions. In fact,
{\crz this restriction of the problem can be written as}
$$
\textrm{(\RP)} \qquad z(S) =\min\left\{\sum_{i\in S} f_i + b_i
x_i^2: \sum_{i\in S} x_i = 1; \ x_i\in\mathbb{R}_+, \ i\in
S\right\}.
$$
{\crz Problem} \RP\  is a convex optimization problem. Define the
following Lagrangean function (without loss of generality, the
constant term $\sum_{i\in S}{f_{i}}$ has been omitted in the
objective function below):
$$
L(x,\mu,\lambda) = \sum_{i \in S}{b_{i}x_{i}^{2}} +
\lambda\left(1-\sum_{i \in S}{x_{i}} \right) - \sum_{i \in
S}{\mu_{i}x_{i}}
$$
The KKT conditions, for the triple $(x^*,\mu^*,\lambda^*)$, are
\begin{eqnarray*}
\nabla_{x} L(x^*,\lambda^*,\mu^*) & = & 0_{q}\\
\sum_{i \in S}{x_{1}} & = & 1\\
\mu^{*^{T}}x^{*} & = & 0\\
x^* & \geq & 0_{q}\\
\mu^* & \geq & 0_{q}
\end{eqnarray*}
It follows from $x_{i}^{*}>0$ that $\mu_{i}^{*}=0$ for all $i\in S$.
Then,
$$\lambda^{*} = \frac{1}{\sum_{j\in S} \frac{1}{2b_j} }$$ and $$x^*_i = \frac{\lambda^*}{2b_i} = \frac{\frac{1}{2b_i}}{\sum_{j\in
S}{\frac{1}{2b_j}}},\quad i\in S$$ satisfy the KKT conditions and,
therefore, are a global optimum for {\crz problem} \RP .



Although coverage diameters may be computed in a closed-form,
choosing the subset $S\subseteq Q$ of active radars is a {\crz
tedious} computational task. {\crz The branch-and-bound algorithm
described in Section~\ref{sec:bounds&algo}} relies on a dual bound
estimation which is {\crz developed in the next subsection}.

\subsection{Lagrangean relaxation of \P} \label{subsec:lagrrel}
We will use Lagrangean relaxation to obtain a lower bound on
$z^{*}$, the optimal solution value of problem \P . Relaxing
constraints $(c1)$ {\crz of \P\ using} nonnegative Lagrangean
multipliers $\kappa_{i}, i=1,\ldots,|Q|$, we obtain the following
problem:
\begin{center}
\begin{tabular}{cc}
(\LRP) & \begin{tabular}{r l r}
        $z_{LRP}(\kappa)=\min$ &
$ \displaystyle \sum_{i \in Q}{(f_{i}-\kappa_{i})y_{i} +
b_{i}x_{i}^2
+ \kappa_{i}x_{i}}$ &\\
        s.t. & $\displaystyle \sum_{i \in Q}x_{i} = 1$ & $(c5)$\\
        & $x \in \mathbb{R}_{+}^{q}$ & $(c6)$\\
        & $y \in \{0,1\}^{q} $ &$(c7)$\\
        \end{tabular}
\end{tabular}
\end{center}
Problem \LRP , a relaxation of \P\ for any $\kappa\geq 0_q$, is
decomposable since optimal values for the $y_{i}$ variables are
independent of the values of the $x_{i}$ variables. In particular,
{\crz we may choose} the following optimal values for $y$:
$$
y^{*}_{i} = \left\{
\begin{array}{rl}
1 & \mathrm{if}\quad f_{i} < \kappa_{i}\\
0 & \mathrm{if}\quad f_{i} \geq \kappa_{i}\\
\end{array}
\right.\quad \mbox{for all } i\in Q.
$$
The remaining convex program, which depends on the $x$ variables only, is:
\begin{center}
\begin{tabular}{cc}
(\LRP$'$) &
\begin{tabular}{r l r}
$z_{LRP'}(\kappa)=\min$ & $\displaystyle b_{i}x_{i}^2 + \kappa_{i}x_{i}$ & \\
                  s.t. & $\displaystyle \sum_{i \in Q}x_{i} = 1$ & $(c8)$\\
                       & $x \in \mathbb{R}_{+}^{q}$ & $(c9)$\\
\end{tabular}
\end{tabular}
\end{center}
and therefore $z_{LRP}(\kappa) = z_{LRP'}(\kappa) + \sum_{i\in
Q}{(f_{i}-\kappa_{i}) y_i^*}$. In order to solve problem \LRP$'$, we
define the following Lagrangean function, where we use multiplier
$\lambda\in \mathbb{R}$ for constraint $(c8)$ and multipliers
$\mu\in\mathbb{R}_+^q$ for nonnegativity constraints $(c9)$:
$$
L_{\kappa}(\lambda,\mu) = \min \sum_{i\in Q}{(b_{i}x_{i}^{2} +
\kappa_{i}x_{i} - \mu_{i}x_{i}) + \lambda\left(1-\sum_{i\in
Q}{x_{i}}\right)}.
$$
%
%
{\crz Then the} KKT conditions are
\begin{eqnarray}
\nabla_{x_{i}}L_{\kappa}(x^{*},\lambda^{*},\mu^{*}) = 2b_{i}
x_{i}^{*} + \kappa_{i} - \mu_{i}^{*} -
                    \lambda^{*} = 0 & \mbox{ for all } \ i\in Q
\label{eq:KKTRLP'1} \\
\sum_{i\in Q}{x_{i}^{*}} = 1 &   \label{eq:KKTRLP'2}\\
\mu_{i}^{*^{T}}x_{i}^{*} = 0 & \mbox{ for all } \ i\in Q
\label{eq:KKTRLP'3}\\
x_{i}^{*} \geq 0, \ \mu_{i}^{*} \geq 0 & \mbox{ for all } \ i\in Q.
\label{eq:KKTRLP'4}
\end{eqnarray}
From~\eqref{eq:KKTRLP'1} we have
\begin{eqnarray}\label{eq:KKTRLP'5}
x_{i}^{*} = \frac{1}{2b_{i}}(\lambda^{*} + \mu_{i}^{*}
-\kappa_{i}^{*}) & \mbox{ for all } \ i\in Q.
\end{eqnarray}
In order to find values for $x_{i}^{*},\lambda^{*},\mu^{*}$ that
satisfy KKT conditions~\eqref{eq:KKTRLP'1}--\eqref{eq:KKTRLP'4}, let
$S$ (so far unknown) include the set of indices that correspond to
positive covering diameters in the optimal solution, {\crz that is}
$S = \{i\in Q : x_{i}^{*} > 0\}$. Given $S$, we have
from~\eqref{eq:KKTRLP'3} that $\mu_{i}^{*}=0$ for all $i\in S$ and
we obtain the following relations:
\begin{eqnarray}
x_{i}^{*} = \frac{1}{2b_{i}}(\lambda^{*} - \kappa_{i}),& \mu_{i}^{*} = 0 &
\mbox{ for all }  i\in S \label{eq:KKTRLP'6}\\
x_{i}^{*} = 0,& \mu_{i}^{*} = \kappa_{i} - \lambda^{*} & \mbox{ for
all } i\in Q\setminus S. \label{eq:KKTRLP'7}
\end{eqnarray}
Furthermore, from~\eqref{eq:KKTRLP'6} and \eqref{eq:KKTRLP'7}, we have that
\begin{eqnarray}
x_{i}^{*} > 0 \Rightarrow \lambda^{*} > \kappa_{i}
& \mbox{ for all } \ i\in S \label{eq:KKTRLP'8}\\
\mu_{i}^{*} \geq 0 \Rightarrow \lambda^{*} \leq \kappa_{i} &  \mbox{
for all } \ i \in Q\setminus S.\label{eq:KKTRLP'9}
\end{eqnarray}
Suppose now, without loss of generality, that the
$\kappa_{i}$ values are in nondecreasing order. From relations
\eqref{eq:KKTRLP'8} and \eqref{eq:KKTRLP'9}, we have:
\begin{equation}\label{eq:KKTRLP'10}
\underbrace{\kappa_{1}\leq\kappa_{2}\leq\ldots\leq\kappa_{h}}_{S}
<\lambda^{*}\leq
\underbrace{\kappa_{h+1}\leq\ldots\leq\kappa_{q}}_{Q\setminus S}.
\end{equation}
Hence, $S$ has the form
$S=\{1,\ldots,h^*\}$, $h^* \leq q$,
%
%
and we obtain the following expression for $\lambda^{*}$ ({\crz
using} equations (\ref{eq:KKTRLP'2}), (\ref{eq:KKTRLP'6}),
(\ref{eq:KKTRLP'7})):
\begin{equation}\label{eq:lambdaott}
\lambda^{*} = \frac{\displaystyle
1+\sum_{i=1}^{h^*}{\frac{\kappa_{i}}{2b_{i}}}}{\displaystyle
\sum_{i=1}^{h^*}{\frac{1}{2b_{i}}}} > 0.
\end{equation}

Since the feasible region of \LRP$'$ is a closed convex set and its
objective function is convex, this problem admits a (finite) optimal
solution. In particular any local optimum that satisfies the KKT
conditions is an optimal solution for \LRP$'$ and vice versa.
Therefore, there must exist an optimal solution $x^*$ of \LRP$'$,
together with corresponding optimal multipliers $\lambda^*\in\Re$,
$\mu\in\Re_+^q$, that satisfy the KKT conditions
\eqref{eq:KKTRLP'1}--\eqref{eq:KKTRLP'4}.

Therefore, once the $\kappa_i$ are arranged in nondecreasing order,
a set of indices $S =\{1,\ldots, h\}$ ($1\leq h\leq q$) necessarily
exists such that \eqref{eq:KKTRLP'10} is satisfied, and expressions
\eqref{eq:KKTRLP'6}, \eqref{eq:KKTRLP'7}, and \eqref{eq:lambdaott},
return an optimal solution $(x^*, \lambda^*, \mu^*)$ to \LRP$'$. 

We may find $h$, that is, the set $S=\{1,\ldots,h\}$ of indices
corresponding to {\crz selected discs,  with a $O(\log{q})$ binary
search}. Because it is the sum of at most $q$ elements, it is
possible to compute $\lambda^{*}$ by~\eqref{eq:lambdaott} in $O(q)$
time. The same time is required to compute the values $x^*_{i}$,
which are at most $q$ and each is computable in constant time ({\crz
using} expression~\eqref{eq:KKTRLP'6}). Additional $O(q)$ steps are
necessary to determine the value for the $y$ variables. {\crz Hence}
the following proposition holds:
\begin{proposition}\label{prop:comptime}
Given a set of $q$ nonnegative Lagrangean multipliers $\kappa \in
\mathbb{R}^{q}_{+}$, the solution of the Lagrangean problem \LRP$'$
can be found in time $O(q\log{q})$, which is the {\crz
computational} cost of ordering the multipliers. $\blacksquare$
\end{proposition}
Recalling that  $z_{LRP}(\kappa) = z_{LRP'}(\kappa) + \sum_{i\in
Q}{(f_{i}-\kappa_{i}) y_i^*}$, we observe that a solution of problem \LRP---and therefore a lower bound for the optimal solution value of \P---can be found in time  $O(q\log{q})$.

\section{An exact algorithm for \P}\label{sec:bounds&algo}
In this section, we develop a branch-and-bound algorithm that finds
an optimal solution of \P . This {\crz implicit} enumeration scheme
exploits the lower bounds (LB) obtained by a subgradient
optimization algorithm {\crz described} in Section~\ref{subsec_LB}
and upper bounds (UB) by an efficient local search-based heuristic
(described in Section~\ref{subsec:UB}). Branching and subproblem
solution strategies are discussed in Section~\ref{subsec:B&B}.

\subsection{
Lower bound via a subgradient algorithm}\label{subsec_LB}
\begin{figure}[h!]
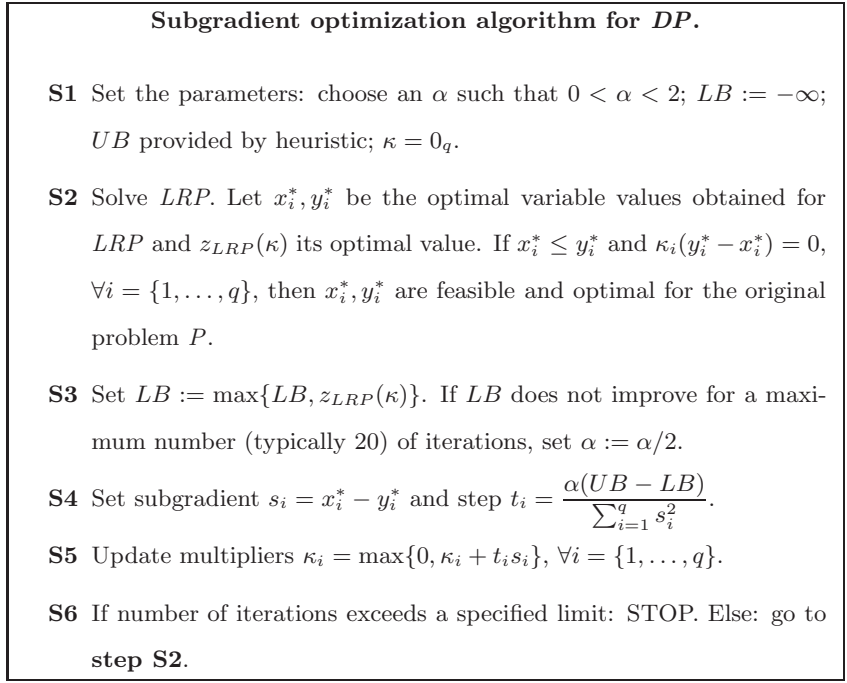

\begin{center}
\fbox{
\begin{minipage}[t]{300pt}\small
\begin{center}
\textbf{Subgradient optimization algorithm for \DP .}
\end{center}
\begin{enumerate}
\item[\textbf{S1}] Set the parameters: choose an $\alpha$ such that $0 <
\alpha <2$; $LB := -\infty$; $UB$ provided by heuristic; $\kappa =
0_{q}$.
\item[\textbf{S2}] Solve \LRP . Let $x_{i}^{*},y_{i}^{*}$ be
the optimal variable values obtained for \LRP\ and $z_{LRP}(\kappa)$
its optimal value. If $x_{i}^{*}\leq y_{i}^{*}$ and
$\kappa_{i}(y_{i}^{*}-x_{i}^{*})=0$, $\forall i =\{1,\ldots,q\}$,
then $x_{i}^{*},y_{i}^{*}$ are feasible and optimal for the original
problem \P .
\item[\textbf{S3}] Set $LB:= \max\{LB, z_{LRP}(\kappa)\}$. If $LB$ does not improve for a maximum number (typically 20) of
iterations, set $\alpha := \alpha /2$.
\item[\textbf{S4}] Set subgradient $s_{i} = x_{i}^{*}-y_{i}^{*}$
and step $\displaystyle t_{i} =
\frac{\alpha(UB-LB)}{\sum_{i=1}^{q}{s_{i}^{2}}}$.
\item[\textbf{S5}] Update multipliers $\kappa_{i}=\max\{0,
\kappa_{i}+t_{i}s_{i}\}$, $\forall i = \{1,\ldots,q\}$.
\item[\textbf{S6}] If number of iterations exceeds a specified limit: STOP.
Else: go to \textbf{step S2}.
\end{enumerate}
\end{minipage}
} 
\end{center}
\caption{\small Subgradient algorithm for \DP .}
\label{fig:subgradalgo}
\end{figure}

For any $\kappa\in\mathbb{R}_{+}^{q}$, the optimal solution value
$z_{LRP}(\kappa)$ of \LRP\ provides a lower bound on the value of
the optimal solution value of \P. We are now interested in
obtaining the best (largest) lower bound by solving the following
\emph{Lagrangean Dual Problem}:
$$
\textrm{(\DP)}\qquad z^{*}_{LRP} = z_{LRP}(\kappa^{*}) =
\max{\{z_{LRP}(\kappa):\kappa\in\mathbb{R}_{+}^{q}\}}
$$
In our approach, the solution of \DP\ is obtained by a standard
subgradient optimization algorithm that is summarized in
Figure~\ref{fig:subgradalgo}. (The actual values of parameters
$\alpha$ and $t_i$ used in the implementation are reported later.)
\normalsize
%
The proposed Lagrangean relaxation method not only provides the
lower bounds that we use in our enumeration scheme but it is also
exploited in an efficient heuristic procedure which is presented in the next section.

\subsection{
Upper bound via a heuristic algorithm}\label{subsec:UB} The basic
idea for this heuristic algorithm based on the Lagrangean Relaxation
is to obtain a feasible solution of \LRP\ by {\crz establishing all
the discs (radars) corresponding to} $x_{i}>0$ (i.e., $i\in S$) and,
possibly, {\crz removing unused discs (i.e., switching off all the
unnecessary radars)} ($i\notin S$).

In general, given a subset $S\subseteq Q$, we may easily compute
feasible values for the coverage diameters $x_{i}$, for all $i\in
S$, using the KKT conditions---as {\crz described} in
Section~\ref{sec:model}. Note that the values $x_{i}$ provided by
the solution of \LRP$'$ are feasible but, in general, they may not
be optimal since the corresponding set of {\crz discs} $S$ may not
be optimal. We may further refine the set $S$ using a simple local
search which exploits KKT conditions to find the cheapest location
and coverage for a given set of {\crz selected discs}.
Figure~\ref{fig:euristicalgo} summarizes the heuristic.

\begin{figure}[hbtp]
\begin{center}
\fbox{
\begin{minipage}[t]{300pt}\small
\begin{center}\textbf{Heuristic algorithm for \P}.\end{center}
\begin{enumerate}
\item[\textbf{H1}] Choose nonnegative values for penalties (Lagrangean multipliers)
      $\kappa_{i}, i=\{1,\ldots,q\}$ (e.g., those found by the
      subgradient algorithm in Figure~\ref{fig:subgradalgo}).
\item[\textbf{H2}] Sort vector $\kappa$ in nondecreasing order.
\item[\textbf{H3}] Compute the set $S$ of {\crz selected discs}
      and the corresponding values for the $x_{i}$ using the KKT-based method
      presented end of Section~\ref{subsec:lagrrel}.
Then, the cost of this
      feasible solution is $z(S) = \min\sum_{i \in S}\left\{f_{i} +
      b_{i}\frac{1/2b_{i}}{\sum_{j\in S}{1/2b_{j}}} \right\} $.
\item[\textbf{H4}] Perform a local [greedy] search on $S$ for a limited number of
      iterations, by
      \begin{enumerate}
      \item Trying to {\crz remove a disc} (starting from the one with the
            largest $f_{i}$ among those {\crz selected}) and computing the
            resultant $z(S)$. Update $S$ if the solution is improved.
      \item Trying to {\crz establish a disc} (starting from the one with the
            smallest $f_{i}$ among those not {\crz located}) and computing the
            resultant $z(S)$. Update $S$ if the solution is improved.
\end{enumerate}
\end{enumerate}
\end{minipage}
} 
\end{center}
\caption{\small Heuristic procedure for \P .}
\label{fig:euristicalgo}
\end{figure}
\normalsize

\subsection{Exact branch-and-bound algorithm}\label{subsec:B&B}
In this section, we present a branch-and-bound algorithm for \P\
that uses the lower and upper bounds developed {\crz above} in
Sections~\ref{subsec_LB} and \ref{subsec:UB}.

\subsubsection{Solution strategy}\label{subsubsec:B&Bsolstrategy}
In the {\crz branch-and-bound} tree, each node $\nu$ represents a
subproblem that is defined by $(i)$ {\crz a set $T$ of selected
discs (i.e. active radars which must be ON)} in the solution, that
is $T = \{i:y_{i} =1; i=\{1,\ldots,q\}\}\subseteq Q$, $(ii)$ a set
of {\crz discs that cannot be in the solution (i.e. radars that must
be OFF)} and $(iii)$ a set of {\crz discs that are not yet decided
upon (i.e., radars that are not yet fixed to ON or OFF)}. If a radar
is OFF, we consider {\crz the corresponding disc} \emph{deleted}
from the set of available {\crz discs (radars)} for that specific
subproblem. {\crz Let $Q(\nu)$ be} the set of available {\crz discs}
at node (subproblem) $\nu$. {\crz Then the} generic subproblem may
be formulated as follows:

\begin{center}
\begin{tabular}{cc}
$(\textrm{\P}(T,\nu))$  & \begin{tabular}{r l l }
       $\min$ & $\displaystyle \sum_{i \in Q(\nu)\setminus
       T}{\left(f_{i}y_{i} + b_{i}x_{i}^{2}\right)} + \sum_{i\in
       T}{\left(f_{i} + b_{i}x_{i}^{2}\right)}$ &\\
         s.t. & $\displaystyle x_{i}\leq y_{i}\quad$  for all $i\in
         Q(\nu)\setminus T$ & $(c1_{\nu})$\\
            & $\displaystyle \sum_{i \in Q(\nu)}{x_{i}}=1$  & $(c2_\nu)$\\
            & $x_{i}\geq 0\quad$  for all $i\in Q(\nu)$ & $(c3_{\nu})$\\
            & $y_{i} \in \{0,1\}\quad$  for all $i\in Q(\nu)\setminus T $ & $(c4_{\nu})$\\
            \end{tabular}
\end{tabular}
\end{center}

Again, similarly to what was done in Section~\ref{subsec:lagrrel}
for problem \LRP , relaxing constraints $(c1_{\nu})$ in a Lagrangean
fashion, using multipliers $\kappa_{i}$ with $\kappa_{i}=0$ for all
$i\in T$, we obtain problem \LRP$(T,\nu)$:
$$
\min \left\{\sum_{i\in Q(\nu)}{\left(b_{i}x_{i}^{2} +
\kappa_{i}x_{i}\right)} + \sum_{i\in Q(\nu)\setminus
T}{\left(f_{i}-\kappa_{i}\right)y_{i}} + \sum_{i\in T}{f_{i}} \ : \
(c2_\nu), (c3_{\nu}) \mbox{ and } (c4_{\nu}) \right\}
$$
which, in turn, is equivalent to
$$
\min \left\{\sum_{i\in Q(\nu)}{\left(b_{i}x_{i}^{2} +
\kappa_{i}x_{i}\right)} \ :\ (c2_\nu), (c3_{\nu})\right\} + \sum_{i\in
Q(\nu)\setminus T,\ f_i < \kappa_i}{\left(f_{i} - \kappa_{i}\right)}
+ \sum_{i\in T}{f_{i}}
$$

Neglecting the last two constant summations, we have a problem in
the $x$ variables which is a special instance of \LRP$'$ defined in
Section~\ref{subsec:lagrrel}. Thus, a lower bound can be computed at
each node by solving the Lagrangean dual of the corresponding
problem \LRP$(T,\nu)$, by means of the procedure summarized in
Figure~\ref{fig:subgradalgo}.

An upper bound at the \emph{root node} is provided by the heuristic in
Figure~\ref{fig:euristicalgo}.

\subsubsection{Branching strategy}\label{subsubsec:enumstrategy}
At node $\nu$ of the enumeration tree, we branch on a binary
variable $y_{i}$, $i\in Q(\nu)\setminus T$, splitting subproblem
$\nu$ into two new subproblems $\nu'$ and $\nu''$. {\crz Disc $i$ is
selected} in $\nu'$ (i.e. $y_{i}=1$ and $T:= T\cup \{i\}$) and it is
deleted in $\nu''$ (i.e. $y_{i}=0$ and $Q(\nu''):=
Q(\nu)\setminus\{i\}$).

Let $\kappa_{i}^{*}$ be the optimal values for the multipliers in
the solution of the Lagrangean dual, and $x^{*}_{i}$, $y^{*}_{i}$,
$i \in Q(\nu)\setminus T$ the optimal variable values obtained for
\LRP$(T,\nu)$. The branching rule is to branch on a variable $y_{i}$, such that $y_{i}^{*}=0$ and
$x_{i}^{*}>0$. If such a variable does not exist (i.e., $x^{*}$ and
$y^{*}$ are feasible for the subproblem $\nu$) then branch on variable $y^*_{i}$, such that $y^{*}_{i} = 1$, $x^*_{i} < 1$ and
$\kappa^*_{i} >0$. If such a variable does not exist then
$\kappa_{i}^{*}(y^{*}_{i}-x_{i}^{*})=0$, for all $i\in
Q(\nu)\setminus T$ and, therefore, the $x^{*}$ and $y^{*}$ are
(feasible and) \emph{optimal} for subproblem $\nu$. The
corresponding node in the enumeration tree is then fathomed.

%

\section{Computational experiments}\label{sec:exp}
The design of the computational experiments is described in the next
subsection while the computational results are discussed in
Section~\ref{subsec:results-and-analysis}. All the results reported
in this section refer to tests performed on a 3.00~GHz Pentium IV,
1024 MB RAM, running Windows XP. The algorithms have been coded in
C++. See~\cite{grandethesis} for more details.

\subsection{Design of experiments}\label{subsec:doe}
Any instance of Problem \P\ is characterized by a pair of vectors
with $q$ components $(b,f)$, representing {\crz discs'} variable and
fixed costs.

We say that a {\crz disc} $i$ \emph{dominates} {\crz disc} $j$ if
$(b_{i} \leq b_{j})$, $(f_{i}\leq f_{j})$, and $(b_i,f_i)\neq
(b_j,f_j)$. In our experiments no {\crz disc} pair exists such that
one is dominated by the other, {\crz since there is no sense in
considering dominated disc types in $Q$}.

Therefore, we impose the following cost {\crz relations}:
$$
b_{1}\leq b_{2}\leq\cdots\leq b_{q}\quad \mbox{and}\quad f_{1}\geq
f_{2}\geq\cdots\geq f_{q}.
$$

We start {\crz with a } special class of instances (\emph{``base
class''}) having the following properties:
\begin{itemize}
\item $(b_{i}\neq b_{j})$ and $(f_{i}\neq~f_{j})$, for
all $1\leq i < j\leq q$.
\item On the average, $b_{i+1}\approx b_i + 1$ for all $1\leq i < q$.
\item $b_{i} = f_{q-i+1}$, for all $1\leq i\leq q$ (to exclude dominated cases).
\end{itemize}

We generate all the instances used in the experiments by suitable
modifications of a randomly generated base class instance. In
particular, any instance is identified by the four integers $(q, s,
t, u)$, where
\begin{enumerate}
\item[$q$:] the number of available {\crz discs} which determines the size of the instance.
\item[$s$:] amplification factor by which $b$ of the base class instance is multiplied. For instances with this parameter, on the
      average, $b\approx\{s, 2s, \ldots, qs\}$.
\item[$t$:] the parameter that characterizes the vector of setup
costs $f$, which is
      obtained using $t\geq 1$ as a multiplication factor of the $b$ vector determined {\crz as above}: on the
      average, $f_{q-i+1}\approx tb_i$ {\crz for instances with this parameter}.
\item[$u$:] the parameter that identifies the configuration
for the test instance where a suitable subset of the cost
coefficients $b_i$, or the setup costs $f_i$, or both, have the same
value. 10 configurations were defined and $u$ was labeled
$0,1,\ldots,9$ where $u=0$ defines the base class {\crz where $b_i$
and $f_i$ all have different values}. For example $u=2$ defines the
class where the $f_i$ for the {\crz selected discs} in the optimal
base class solution are set to the maximum $f$ value in the base
class. Parameter $u$ attempts to make systematic changes with
respect to $f_i$ and $b_i$ values in various instances. While some
other $u$ labels are described later in this paper,
see~\cite{grandethesis} for more details on the other $u$ labels.
\end{enumerate}
As an example, the class $(50,1,100,0)$ refers to instances with
$50$ {\crz discs}, all {\crz different types} (since $u=0$), cost
coefficients $b_i$ as in the base class, and setup costs $f_i$
amplified by a factor $t=100$.

{\crz A set} of preliminary tests were performed to determine the
largest instances that our algorithm is able to solve optimally, in
order to design our experiments. Results are reported in
Table~\ref{tab:dimensioning}. All the instances of this preliminary
test-set belong to the class $(q,1,1,0)$. The branch-and-bound
algorithm solved instances up to $q=400$ in less than 16 hours. No
instance with $q=500$ was solved within the same time limit. Almost
all the instances with $q$ up to $200$ are solved within one hour.
\begin{table}[\htbp]
\begin{center}\small
\begin{tabular}{c c c c c c}
\hline\hline $q$ & $150$ & $200$ & $350$ & $400$ & $500$\\
\hline CPU time & 1217.97 s. & 2544.76 s. & $\sim 3.5$ hr. & $\sim 15.0$ hr.& $> 16$ h\\
\hline\hline
\end{tabular}
\caption{\small Preliminary test-set results}
\label{tab:dimensioning}
\end{center}
\end{table}

Based on these preliminary results, we planned our experiments with
the following sizes: $q\in \{10,25,50,\\100,200,350,400\}$. Maximum
running time was set to 1 hour (CPU time), except for the case
$q=400$ where there was no timeout requirements. For each class
except the $q=400$ case, 10 random instances were generated and
%
%
the following \emph{average} quantities were tracked:
\begin{itemize}
\item CPU time.
\item Number of nodes in the enumeration tree.
\item Depth of the enumeration tree.
\item Upper bound at the root node.
\item Optimal solution value, if any. (A minus ``--'' symbol is
shown when the optimum is not reached within the time limit.)
\item Best lower bound available after 1 hour.
\item Percentage gap $(UB - LB)/UB$.
\end{itemize}
%
%

\subsection{Results and analysis}\label{subsec:results-and-analysis}

Table~\ref{tab:riassuntoesp} summarizes the results of the
experiments. Each row shows, in order, the quantities of the above
list, for one class of instances. Class name is reported in the
first column where an asterisk ``$^*$'' denotes that, in at least
one instance of the class, the algorithm did not reach the optimal
solution value within the time limit (i.e. CPU time greater than $3600$ sec.).
The table also gives the initial ``gap'' between the UB and the LB
at the root node, computed as $(UB-LB)/UB$. The results for classes
with $u=3,4,6,8,9$ are not reported in Table~\ref{tab:riassuntoesp}
for the sake of brevity. For these classes, the performance of the
algorithm is indeed comparable or even better than those reported.

\begin{figure}[!htbp]
\begin{center}
\includegraphics[scale=0.6,angle=0]{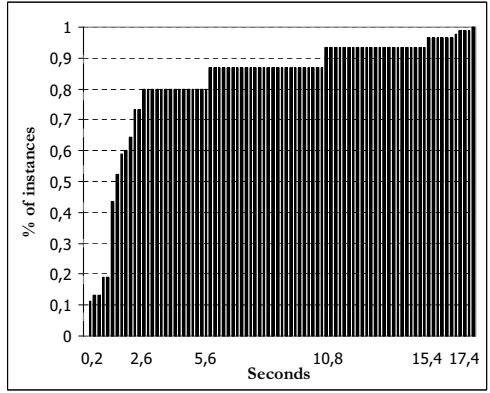}
\end{center}
\caption{\small CPU time cumulative distribution ($q=25$, time in
seconds).} \label{fig:distremp}
\end{figure}

First, we highlight the excellent performance of the heuristic:
\emph{in all the experiments} the value found by this procedure
($UB_{root}$) equals the optimal value (opt.) found by the
branch-and-bound algorithm.
\begin{table}[!htbp]
\begin{center}\footnotesize
\begin{tabular}{c*{7}{c}}
\hline\hline \begin{tabular}{c}instance\\$(q,s,t,u)$\end{tabular} &
\begin{tabular}{c}CPU\\time (s.)\\\end{tabular} & nodes' \#
& depth & $UB_{root}$ & opt. & $LB$ & gap$^\dag$\\
\hline 10,10,1,0 & 1.547 \rule{0pt}{10pt}& 29 & 10 & 77.368 & 77.368
&
44.8017 & 42.01\%\\
10,10,1,1  & 1.469  \rule{0pt}{10pt}& 27 & 10 & 80 & 80 & 46.696 & 41.63\%\\
10,10,1,5 & 1.562 \rule{0pt}{10pt}& 37 & 10 & 110 & 110 & 75.526 & 31.64\%\\
10,1,100,0 & 0.063 \rule{0pt}{10pt}& 0 & 0 & 506 & 506 & 506 & 0\%\\
10,1,100,1 & 0.062 \rule{0pt}{10pt}& 0 & 0 & 110 & 110 & 110 & 0\%\\
10,1,100,5 & 0.047 \rule{0pt}{10pt}& 0 & 0 & 506 & 506 & 506 & 0\%\\
100,10,1,0 & 464.81 \rule{0pt}{10pt}& 987 & 100 & 345.96 & 345.96 & 135.31 &60.89\%\\
100,10,1,1 & 464.25 \rule{0pt}{10pt}& 1009 & 100 & 348.35 & 348.35 & 135.74 & 61.03\%\\
100,10,1,5 & 932.47 \rule{0pt}{10pt}& 1955 & 96 & 496.22 & 496.22 & 172.64 & 65.21\%\\
100,1,100,0 & 31.937 \rule{0pt}{10pt}& 135 & 67 & 200 & 200 & 187.44 & 6.28\%\\
100,1,100,1 & 32.125 \rule{0pt}{10pt}& 131 & 65 & 200 & 200 & 187.5 & 6.25\%\\
100,1,100,5 & 33.047 \rule{0pt}{10pt}& 131 & 65 & 596 & 596 & 584.92 & 1.86\%\\
200,10,1,0 & 2586.01 \rule{0pt}{10pt}& 2737 & 200 & 539.14 & 539.14
& 189.31 &
64.89\%\\
200,10,1,1 & 2866.91 \rule{0pt}{10pt}& 2527 & 200 & 540.8 & 540.8 &
189.66 &
64.93\%\\
200,10,1,5* & $>3600$\rule{0pt}{10pt} & $>2597$ & $\geq 195$ &
737.98 & -- &
227.93 & 69.12\%\\
200,1,100,0 & 208.7 \rule{0pt}{10pt} & 341 & 169 & 300 & 300 & 241.54 & 19.49\%\\
200,1,100,1 & 189.281 \rule{0pt}{10pt} & 361 & 174 & 300 & 300 & 241.67 & 19.45\%\\
200,1,100,5 & 206.38 \rule{0pt}{10pt}& 337 & 167 & 696 & 696 & 639.7 & 8.09\%\\
350,10,1,0* & $>3600$ \rule{0pt}{10pt}& $>2021$ & $\geq 345$ & 775.7
& -- &
234.2 & 69.81\%\\
350,10,1,1 & 959.97 \rule{0pt}{10pt}& 1827 & 345 & 776.91 & 776.91 &
234.74 &
69.79\%\\
350,10,1,5* & $>3600$ \rule{0pt}{10pt}& $\gg 1289$ & $\geq$ 344 &
1022.49 & -- &
285.4 & 72.09\%\\
350,1,100,0 & 985.38 \rule{0pt}{10pt}& 719 & 350 & 450 & 450 & 301.56 & 32.99\%\\
350,1,100,1 & 959.97 \rule{0pt}{10pt}& 703 & 350 & 450 & 450 & 298.51 & 33.67\%\\
350,1,100,5 & 964.44 \rule{0pt}{10pt}& 673 & 328 & 846 & 846 & 700.2 & 17.23\%\\
400,1,1,0 & 36705.2 \rule{0pt}{10pt}& 17373 & 400 & 84.71 & 84.71 & 24.28 & 71.34\%\\
\hline\hline
\end{tabular}\normalsize
\caption{\small Experiments results. ($\dag$ Gap is computed as the
initial $(UB-LB)/UB$, where $LB$ is the (initial) lower bound at the
root node.)} \label{tab:riassuntoesp}
\end{center}
\end{table}
\normalsize
A~few comments are in order:
\begin{enumerate}
\item Not surprisingly, branch-and-bound computational time is
      strictly  related to the
      enumeration tree size: Table~\ref{tab:timenodescorrelation}
      shows strong positive correlation of CPU time with the
      number of nodes in the enumeration tree.
\item For a given number of {\crz discs} $q$, we note that
      the CPU time does depend on the particular cost configuration,
      that is, on the particular pair $(s,t)$. The most difficult
      instances have $t=1$ (i.e., $b_{i}\approx f_{q-i+1}$); and,
      viceversa, the
      larger the $t$ in comparison to $s$, the faster the computation.

\item Classes with $u=2$ and $u=5$ are the hardest. Parameters are
      chosen so that finding the set of {\crz selected
discs}
      becomes more difficult.
      For any instance $I$ of the base class, we build the corresponding
      $u=5$ instance $I'$ as follows. Let $S$ be the set of {\crz selected discs} in
      the optimal solution of $I$. The costs in $I'$ are:
      $b'_i = b_i$ if $i\not\in S$, $b'_i = \max\{b_1,\ldots, b_q\}$
      if $i\in S$. Analogously $f'_i = f_i$ if $i\not\in S$,
      $f'_i = \max\{f_1,\ldots, f_q\}$ if $i\in S$.

      The $u=2$ class is designed similarly but with $b'_i=b_i$, for all $i = 1, \ldots, q$.
      (Experimental results for
      the $u=2$ class {\crz were similar} to
      those of the $u=5$ class; therefore Table~\ref{tab:riassuntoesp} reports only the latter.)
\item
      Figure~\ref{fig:distremp} shows the CPU
      time cumulative distribution. The histogram {\crz was obtained}
      empirically over 90 instances (with $q=25$): 80\% of the instances are solved in a time smaller
      than 2.6 s. while the largest CPU time is an order of magnitude
      higher (17.2 s.).
\end{enumerate}
\begin{table}[!htbp]
\begin{center}\small
\begin{tabular}{c c c c c}
\hline\hline $q$ & 10 & 25 & 50 & 100\\
\hline$\rho_{nodes}$ & 0.982 & 0.998 & 0.999 & 0.999\\
\hline\hline
\end{tabular}
\caption{\small Correlation coefficient between CPU time and number
of nodes of the enumeration tree.} \label{tab:timenodescorrelation}
\end{center}
\end{table}\normalsize
%

We noted previously that fixed costs are related to the choice of
the subset $S$ and they heavily affect the computational effort
required by an instance. An evidence of this fact is illustrated in
Table~\ref{tab:impattocostiFeB} where the results of the experiments
with $q,s,t = 10,1,1$ are
%
%
\begin{table}[!htbp]
\begin{center}\small
\begin{tabular}{c c c}
\hline\hline \begin{tabular}{c}instance\\($q,s,t,u$)\end{tabular} & $S^*$ & \begin{tabular}{c}Opt. radii\\$x_{i}, i\in S^{*}$\end{tabular}\\
\hline 10,1,1,0 & \{9,10\} & \{0.526, 0.474\}\\
 10,1,1,1 & \{9,10\} & \{0.5, 0.5\}\\
10,1,1,3 & \{7,8,9\} & \{0.377, 0.330, 0.293\}\\
\hline\hline
\end{tabular}
\caption{\small Set $S^*$ (optimal set) response to $f_{i}$ and
$b_{i}$ variations for $q=10$ and $u =\{0,1,3\}$.}
\label{tab:impattocostiFeB}
\end{center}
\end{table}\normalsize
compared for three values of $u$ ($u=0,1,3$). Note that the $u=0$ class is the base class.

Given an instance $I$ of the base class, we build the corresponding
      $u=1$ ($u=3$ respectively) instance $I'$ as follows. Let $S$ be the set of {\crz selected discs} in
      the optimal solution of $I$. The costs in $I'$ are:
      $b'_i = \max\{b_1,\ldots, b_q\}$ ($f'_i = \min\{f_1,\ldots, f_q\}$ respectively)  for $i\in S$. All the other parameters remain equal to those in the base class.

When the $b_i$ are varied (compare classes with $u=0$ and $u=1$ in
Table~\ref{tab:impattocostiFeB}) the optimal solutions are slightly
different. However, when the $f_i$ vary (compare classes with $u=0$
and $u=3$) the two optimal solutions drastically differ from each
other: there are three {\crz discs selected} instead of two.

%
\section{Conclusions and future work}\label{sec:concl}
In this paper we introduced and addressed the problem of covering a
single line segment with radar sensors having a circular field of
view. When the sensors are required to have identical radius, a
simple polynomial search solves for optimal radius and number of
sensors. When the sensors are modeled with variable diameter discs
the problems becomes hard. We provided an exact solution algorithm
which is based on a Lagrangean relaxation and a subgradient
algorithm to find a lower bound (see Figure~\ref{fig:subgradalgo}).
A feasible solution is provided by the heuristic summarized in
Figure~\ref{fig:euristicalgo}. These bounds were exploited to design
a branch-and-bound algorithm. Extensive computational testing, based
on approximately 400 experiments, showed that the developed
heuristic performs very well; the upper bounds were always equal to
the optimal solution whenever the latter was known (see
Table~\ref{tab:riassuntoesp}). The experiments also show that setup
costs play a crucial role both in computational effort and
attainment of the optimal solution set.

Several directions for future work are being pursued. The two-arc
network and planar tree network cases are being investigated (both
for single and multiple {\crz discs types}, for fixed and variable
radius). Since {\crz discs} can be located both on the arcs and on
the plane, these are mixed network--planar problems and the
development of locational models and algorithms is indeed very
challenging.

\section*{Acknowledgments}

The authors wish to acknowledge the support of the ATLAS Center at
the University of Arizona where most of this research was conducted
and the support of Waveband Inc. that partially funded this effort
through its SBIR Contract ARMY03-T18 with the US Army.

The authors also gratefully acknowledge
the helpful comments and constructive suggestions
of the two anonymous referees.

\bibliographystyle{abbrv}
\bibliography{COR_biblio}

\end{document}